\documentstyle[11pt,aaspp4]{article}

\begin{document}

\title{{\bf The scale of homogeneity in the Las Campanas Redshift Survey}}
\author{Luca Amendola and Emilia Palladino}
\affil{Osservatorio Astronomico di Roma, \\
Viale  del Parco Mellini 84, 
00136 Roma, Italy}

\begin{abstract}
We analyse the Las Campanas Redshift Survey using the integrated conditional
density (or density of neighbors)  in volume-limited subsamples up
to unprecedented scales (200 Mpc/$h$) in order to determine without
ambiguity the behavior of the density field. We find that the survey is well
described by a fractal up to 20-30 Mpc/$h$, but flattens toward homogeneity
at larger scales. Although the data are still insufficient to establish with
high significance the expected homogeneous behavior, and therefore to rule
out a fractal trend to larger scales, a fit with a CDM-like spectrum with
high normalization well represents the data.
\end{abstract}



Following seminal work of Pietronero and coworkers, (see e.g. 
Pietronero et al. 1996), the possibility of a large-scale
fractal distribution of
the galaxies has been investigated by various authors.
In the current literature, however,
 there are several conflicting estimates of the
largest scale at which the galaxy distribution can be approximated by a
fractal, ranging from a few Megaparsecs (e.g. Peebles 1993), to  20  
Mpc/$h$
(Davis 1996), to 40 Mpc/$h$ (Cappi et al. 1998), up 
to more than 100 Mpc/$h$ (e.g.
Pietronero et al. 1996). A fractal distribution with dimension $D$ is
characterized by the property that the correlation function 
\begin{equation}
g(r)=1+\xi (r)
\end{equation}
decreases as a power law, $\sim r^{3-D}$. Consequently, the average density $
\rho _c$ of galaxies at distance $r$ from another galaxy , or conditional
density, also decreases as $\sim r^{3-D}$ since, by the definition of
correlation function, $\rho _c=\rho _0[1+\xi (r)],$ where $\rho _0$ is the
cosmic average density.

Naturally, one can infer the scale at which fractality gives way to
homogeneity from several other observations, like the cosmic backgrounds,
although the conclusions are bound to be model dependent. The availability
in recent years of deep redshift surveys allows finally to study the matter
distribution directly from its primary tracers, the galaxies. The deepest
galaxy redshift survey 
so far published is the Las Campanas Redshift Survey (LCRS),
 Schectman et al. (1996). LCRS   contains 23,697 galaxies with 
an average redshift $
z=0.1$ , distributed over six 1.5$^0\times $80$^0$ slices.
In this paper we determine the behavior of $\widehat g(r)$, the volume
integral of $g(r)$, and of the fractal dimension $D(r)$, in volume-limited
subsamples of LCRS up to $r=$200 Mpc/$h$, the largest scale so far investigated
with such a statistics, by making use of purposedly designed cells. This
scale is more than four times the scale previously reached by Cappi et al.
(1998) using SSRS2.



Let us begin by  discussing why the statistics $g(r)$ is particularly
convenient for our purposes.
By far the most popular two-point 
estimators that have been used to
investigate the clustering of the galaxies are
the  correlation function $\xi (r)$ and the power spectrum $P(k)$,
 a Fourier conjugate pair. 
However, the simplest statistics to use
to determine the fractal properties of a distribution is $g(r)$,
 or  $\widehat{g}(r)=V^{-1}\int_0^r g(r') dV$, in   terms of which a fractal is defined. 
There is also another reason to use $\widehat{g}(r)$ to study the galaxy
clustering. The spectrum of a finite survey is actually the convolution of
the true power spectrum with the survey geometry. As a consequence, both the
shape and the amplitude of the estimated spectrum are different from the
true spectrum. In particular, the estimation of the average density from the
sample itself forces the spectrum to vanish for $k\rightarrow 0$, so that
the detection of a turnaround in the spectrum is often suspect (see for
instance Sylos Labini \& Amendola 1996). A similar problem occurs with the
correlation function: the integral constraint (Peebles 1980) forces the
correlation function to become negative at some scale, distorting its shape.
In fact, using a subscript $s$ to denote quantities estimated in a finite
sample of size $R_s$, and employing the fact that the density $\rho _s$ in a
sample around an observer is a conditional density, the following relation
holds 
\begin{equation}
\xi _s(r)=\frac{\xi (r)-\widehat{\xi }(R_s)}{1+\widehat{\xi }(R_s)}\,,
\end{equation}
which shows explicitely the integral constraint $\int \xi _sdV_s=0.$ This
problem is clearly absent from the statistics $g(r)$ because $g_s$ and $g$
are simply proportional, 
$g_s(r)=g(r)/\widehat{g}(R_s)\,.$  
It follows that, contrary to what happens for $\xi $ and $P(k)$, the slope
of $g_s$ is an unbiased estimator of the slope of the clustering trend, i.e.
of its fractal dimension. In fact, we can evaluate the fractal dimension as 
\begin{equation}
D=3+\frac{d\log \widehat{g}}{d\log r} \,,
\end{equation}
and verify  that $D_s=D$. Therefore, to test the claim
of fractality it is necessary to check whether $g(r)$ can be approximated by
a power-law, in which range of scales, and with which slope. The approach to
homogeneity is then characterized by  $D\to 3$, that is, to a
flattening of $g(r).$

The statistics $g(r)$ is a differential quantity. When applied to surveys
with a relatively small density, as are the volume-limited slices of Las
Campanas, it tends therefore to be very noisy. In this case, one can smear
out the noise integrating over cells, obtaining the integrated correlation $
\widehat{g}(r)$. Here an important problem arises. Consider a distribution
of particles described by a statistically isotropic correlation function $
\xi (r)$. When we evaluate the expected total number of neighbors within a
distance $r$, we are performing the integral 
\begin{equation}
\widehat{\xi }(r)=\frac 3{r^3}\int_0^r\xi (r')r'^2dr' \,. \label{defrad}
\end{equation}
If the cell  is not spherical, the integral becomes 
$
\widehat{\xi }=\int \xi (r')W(r',\theta ,\phi )dV
$,
where the window function $W(r,\theta ,\phi )$ is defined to be constant
inside the cell and zero outside and  is normalized to unity. Of course,
only the part of the cell completely contained within the survey has to be
considered. However, if the cells are not spherically symmetric, the value
of $\widehat\xi (r)$ depends on the exact form of $W(r,\theta
,\phi )$, and in general is different from the definition in (\ref{defrad}).
Moreover, $W(r,\theta
,\phi )$ may vary from cell to cell. The obvious solution to this
problem is to restrict the analysis to spherical cells. However, this 
limits the scale to the largest sphere contained within the survey
boundaries which, for most surveys, can be very small: e.g., less than 
10 Mpc/$h$ by radius for LCRS.
As a matter
of fact,
so far all the works who used the $\widehat{g}$ statistics adopted spherical
cells, thereby limiting the scales to less than 50 Mpc/$h$. Since at this
scale the fractal behavior is more or less within the standard description
(e.g., CDM), it is crucial to extend the analysis to larger scales. The
simplest way to do so is to consider radial cells, that is, cells whose
window function can be factorized in a radial and an angular function, both
normalized to unity 
\begin{equation}
W(r,\theta ,\phi )=W_r(r)W_\Omega (\theta ,\phi )\,.
\end{equation}
In this case in fact the function $\widehat{\xi }(r)$ is the same as in
spherical cells, since the angular factor can be integrated out. The
advantage is that one can design a radial cell such as to maximize the scale 
$r$, still fitting it within the survey. As elementary as it is, this method
has never been used in the literature on the galaxy clustering.

Before going to the data analysis, let us estimate the variance of $\widehat g$.
Let us first assume that the three-point correlation function can be written
 as (Peebles 1980) 
$
\varsigma _{ijk}=Q[\xi _{ij}\xi _{jk}+\xi _{ij}\xi _{ik}+\xi _{ik}\xi _{jk}]
$,
where $Q$ is independent of the spatial coordinates. Then, the variance of $
\widehat{g}$ evaluated in $N_c$ independent cells 
 containing on average $N_0$
galaxies is (Peebles 1980; Amendola 1998) 
\begin{equation}
\sigma _{\widehat{g}}^2\equiv N_c^{-1}\left[ N_0^{-1}(1+\widehat{\xi }%
)+\sigma ^2-\widehat{\xi }^2+Q(\widehat{\xi }^2+2K_2)\right]\,,  \label{varcd}
\end{equation}
where $\sigma ^2=\int W_1dV_1W_2dV_2\xi _{12}$ . The first term in Eq. (\ref
{varcd}) is the Poisson noise.
Inserting the power spectrum we have 
\begin{equation}
\sigma ^2=(2\pi ^2)^{-1}\int P(k)W_c(k)k^2dk\,,  \label{sigma}
\end{equation}
where $W_c(k)$ equals $W(k)^2$ for {\it spherical cells}, but
 has to be evaluated numerically in the more 
 general case that we study here (see
Amendola 1998). The last term in the expression (\ref{varcd}) can be written
as (Peebles 1980) 
$
K_2=\int W_1dV_1W_2dV_2\xi (r_1)\xi (r_{12})
$.
In the important case in which $\xi $ is a power law, it can be shown that a
very good approximation is $K_2\simeq \sigma ^2\widehat{\xi }$. For
instance, if $\xi \sim r^{-1}$, it turns out that $K_2=1.04\sigma ^2
\widehat{\xi }$. In the following, we will always approximate $K_2$ in this
way.

Another problem  arises in practice, namely that the cells we use
are not independent, both because they 
are partially overlapping and because the
clustering scale may be larger than the distance from cell to cell. The
effect of the correlation is to reduce the number $N_c$ at the denominator
in Eq. (\ref{varcd}). For instance, if the cells oversample the volume by a
factor of two, it means that a cell out of two is redundant, and the
effective number of cells can be taken as $N_c/2$. In general, the number of
effective independent cells may be approximated as 
$
N_e=\min (N_c,V/V_c) 
$,
where $V_c$ is the cell volume, although of course even this is an
overestimation of the independent cells . Naturally, we could use $N-$body
simulations to estimate the errors including the cell-to-cell correlation,
but then we should generate a different simulation for any model we want to
compare with; moreover, any finite $N-$body will inevitably cut large
scale power, which in the case of testing fractals is a particularly severe
limitation. Some comparisons with $N-$body shows that Eq. (\ref{varcd})
underestimates the errors only by 30\% at most.

In the case of an exact fractal, the
expected value of $\widehat{g}$ inside a spherical region of radius $r$
embedded in a larger box of size $R_0$ is (Coleman \& Pietronero 1992)
\begin{equation}
\widehat{g}(r)=(r/R_0)^{D-3} \,. \label{densint}
\end{equation}
It is not difficult to show that, neglecting $\sigma ^2$ and $
\widehat{\xi }$ with respect to $\widehat{\xi }^2$ , i.e. in the limit of $
R_0\gg r$, and neglecting the Poisson noise, the variance in a fractal is 
\begin{equation}
\sigma _{\widehat{g}}^2\simeq N_c^{-1}\left[ Q\left( 1+\frac{2DJ_2(\gamma )}
3\right) -1\right] \widehat{g}^2 \,, \label{fracvar}
\end{equation}
where $\gamma=3-D$, and 
$J_2(\gamma )=72/[(3-\gamma )(4-\gamma )(6-\gamma )2^\gamma ]$. Notice
that the relative error on $\widehat{g}$ is independent on the scale, as
indeed is found numerically (Amici \& Montuori 1998). For instance, for $Q=1$
and $D=2$, as some observations suggest, $\sigma _{\widehat{g}
}^2=N_c^{-1}(8/5)\widehat{g}^2$.

An important consequence of Eq. (\ref{fracvar}) is that the relative error
of the conditional density measured in a single cell can be very large for a
fractal, more than 100\%. Then, the average density in a sample of galaxies
around us,  a conditional density, has such a large variance, in a
fractal, that it gives in practice no information.
 A further consequence is that the variance of the
amplitude of the correlation function, and of related quantities as $r_0$
and $\sigma _8$, makes the use of the correlation {\it amplitude}, 
as opposed to its slope, useless in the
case of fractals. This problem applies also to the case in which the density
of a sample $n(r)$ is measured as a function of the distance from the
observer, without averaging over several
cells. 

If we define the scale of homogeneity as the scale at which $\widehat{g}$
flattens so that $D\ge 2.9$,
then we can quantify it  in any given CDM model.
Clearly, this scale will be larger the
higher is the normalization $\sigma _8$. It turns out that if $\sigma
_8\approx 1.5$, as observed for bright galaxies in SSRS2 (Benoist et al
1996) the CDM homogeneity scale can be as large as $50$ Mpc/$h$, and reach 
$\approx 70$ Mpc/$h$
for clusters ($\sigma_8\approx 2$): therefore,   only at
 scales larger than 50 Mpc/$h$ the
gap between the pure fractal model and the standard scenarios begins to be
significant.


The $\widehat{g}$ statistics in spherical cells has been applied to several
galaxy surveys (Pietronero, Montuori \& Sylos Labini 1997, Sylos Labini and
Montuori 1997, Sylos Labini et al. 1998, Cappi et al. 1998). Here we
summarize only the results from the deepest of such surveys, SSRS2 (Da Costa
et al. 1994) .
SSRS2 includes 3600 galaxies in 1.13 sr of the southern sky, down to an
apparent magnitude of 15.5. The results of the analysis in Cappi et al.
(1998) indicate that the conditional density decreases as $r^{-1}$ from 1 to
40 Mpc/$h$ in all the volume limited
 samples considered, implying fractality with $D=2$
on these scales, with no indication of flattening. Similar 
results are presented
in Sylos Labini and Montuori (1997) for the APM-Stromlo survey and in Sylos
Labini et al. (1998) for  CfA2 and SSRS2 surveys. These are the
largest scales that can be probed via this method. In
Cappi et al. (1998) the conditional density
from the observer alone, i.e. from the vertex of the sample, has also been
studied, in order to extend the range. The results are that the sample is
fractal up to 40-50 Mpc/$h$ and tends to flatten above this scale. However,
the errors, as expected, are quite large; within 2$\sigma $ the deepest
samples include all values between $D=2.7$ and $D=4$. The errors quoted in
Cappi et al. (1998), moreover, do not include the ensemble variance that,
for fractals, is very large, as already mentioned.

We applied our method of the radial cells to the Las Campanas redshift
survey, LCRS, the deepest redshift survey so far studied. LCRS contains
fields which includes galaxies with magnitude between 16.0 and 17.3 , and
fields with limits 15.0 and 17.7. Every field $i$ has associated a sampling
factor $0<f_i<1$ which is the fraction of the galaxies randomly chosen out of
the total number in the field within the magnitude limits; the weights $1/f_i$
 must be taken into account in the statistics.
We considered the only slice with all fields of large magnitude range (slice
at -12$^0$). We evaluated the conditional density integrated in radial cells
with a shape and orientation such as not to intersect the survey boundaries.
We cut the sample into four VL subsamples,
 denoted as VL147-297, VL 190-330, VL280-410, and VL224-437 (see Fig. 1 and
Table I)
where the numbers give the lower and upper cutoff distance 
(the two cuts are necessary because LCRS has two limiting magnitudes).
 The use of volume-limited samples avoids the uncertainties
connected to the  radial selection functions. The correlation function
in the full magnitude-limited sample is evaluated in Tucker et al. (1997).
The results for $\widehat{g}(r)$ are shown in Fig. 2. The errors are
calculated from Eq. (\ref{varcd}) using two models, the standard CDM with 
galaxy normalization $
\sigma _{8g}=1.5$ and $\Gamma =0.3$ , and the pure fractal with $D=2$ and $
Q=1$. For comparison, we evaluated the errors from a simulation of
a standard CDM model, and found that our theoretical estimate of the errors
is approximate to better than 30\%. It can be seen that there is an approximate $
D\approx 2 $ fractality on small scales, up to 20 or 30 Mpc/$h,$ just as in the
SSRS2 case, followed by a flattening of the slope. 
The results from SSRS2 of Cappi et al. (1998;
sample with magnitude cut at
 -20) obtained with full spheres 
 are reported in Fig. 2 along with VL147-297,
the LCRS sample with average magnitude
closer to SSRS2. 
As we can see, the scales we reach here
are the {\it largest scales ever reached} for the $\widehat{g}(r)$
statistics. In the case of VL280-410,
the trend is decreasing, albeit with a change in slope, down to more than
100 Mpc/$h$, while for VL224-437 (the most sparse sample) a very noisy
flattening is reached already at 40 Mpc/$h.$ In Fig. 3 we plot the fractal
dimension $D$ as a function of $r,$ evaluated as least square slope of sets
of five contiguous points of $\widehat{g}(r).$ The tendence to $D=3$ is very
clear. It is also seen that the dimension is not really constant in any
range, although the scatter is too large to infer a clear trend.

From these results we can conclude that there is a tendency to
homogenization around 50-100 Mpc/$h$, as expected from a CDM model. However,
we remark that we did not detect a clear homogeneous behavior, that is $
\widehat{g}(r)=1,$ not even at more than 100 Mpc/$h$, except for the most
sparse sample. This leaves again space for a fractal behavior to larger
scales, especially in view of the large errors associated to a fractal. The
dimension would be however closer to 3 at large scales. 

We compared our results to the CDM-like power spectrum 
\begin{equation}
P(k)=AkT^2(\Gamma ,k)G(\Omega _m,\Omega _{\Lambda},
\sigma _{8g},\sigma _{8m},\sigma _v) \,,
\end{equation}
where $T^2(\Gamma ,k)$ is the transfer function of Bond \& Efstathiou (1984)
, $G$ includes the redshift correction of Peacock \& Dodds (1994)
and the non-linear correction of Peacock \& Dodds (1996), $\sigma _v$ is
the line-of-sight velocity dispersion, and the subscripts $\Lambda,g$ and $m$
refer to the cosmological constant, the
galaxies and  the total matter, respectively. 
In Fig. 2 we compare LCRS with a tentative  model with $\Gamma =0.24$, $
\sigma _{8g}=1.4$, assuming , for the other parameters, the values $
\sigma _{8m}=1$ , $\Omega _{tot}=1$ ,$\Omega _m=0.4,\Omega _\Lambda
=0.6, h=0.6$ , $\sigma _v=300$ km/sec.
Finally, we find the best fit to $\widehat{g}$ by varying $\Gamma $ and $
\sigma _{8g}$ for scales larger than 30 Mpc$/h$, in order to avoid the
non-linear corrections at smaller scales. The results are listed in
Table I. The average for all four samples is $\sigma _{8g}=1.5$, $\Gamma
=0.3$.

To summarize, in this Letter we have applied the technique of radial cells
to LCRS, pushing the analysis of fractality to 200 Mpc/$h.$ We have shown
that a proper treatment of the errors is important in order to compare 
alternative models, and is particularly crucial for a fractal, in which the
variance tends to be very large. We have also shown that the use of radial
cells allows very large scales to be probed. The conclusion is that LCRS is
well described by a CDM model with a high normalization, up to $\sigma
_8\approx 1.6$ for the deepest, and thus brightest, sample. The trend can be
also approximated as a $D\approx 2$ fractal up to 20-30 Mpc/$h,$ but shows a
clear flattening afterward.

\acknowledgments

We thank Ruth Durrer, Michael Joyce, Marco Montuori and Francesco Sylos
Labini for very interesting discussions throughout the preparation of this
work.

\newpage\

\figcaption
{ Two of the volume limited subsamples of LCRS that we
analyse in this work.}

\figcaption{ The function $\widehat{g}(r)$ \ in various VL samples of\
LCRS with the errors expected in a CDM model and in a fractal $D=2$   
model (dotted lines). The thin line is the tentative CDM fit (see
text). In the upper right panel we also report for comparison
$\hat g(r)$ from a sample
of SSRS2, from Cappi et al. (1998)}

\figcaption{ The function $D(r)$  for the four samples, obtained as the
least square slope of sets of five contiguous values of $\widehat{g}(r)$}

\begin{table*}
\begin{center}
\begin{tabular}{lllll}
\tableline
Sample & $\langle M\rangle $ & $N_{gal}$ & $\Gamma $ & $\sigma _{8gal}$ \\ 
\hline
147-297 & -20.43 & 878 & 0.10 & 1.10 \\ 
190-330 & -20.75 & 1081 & 0.55 & 1.55 \\ 
224-437 & -21.34 & 810 & 0.30 & 1.50 \\ 
280-410 & -21.26 & 793 & 0.40 & 1.65  \\ 
\end{tabular}
\end{center}
\tablenum{I}
\caption{ }
\end{table*}
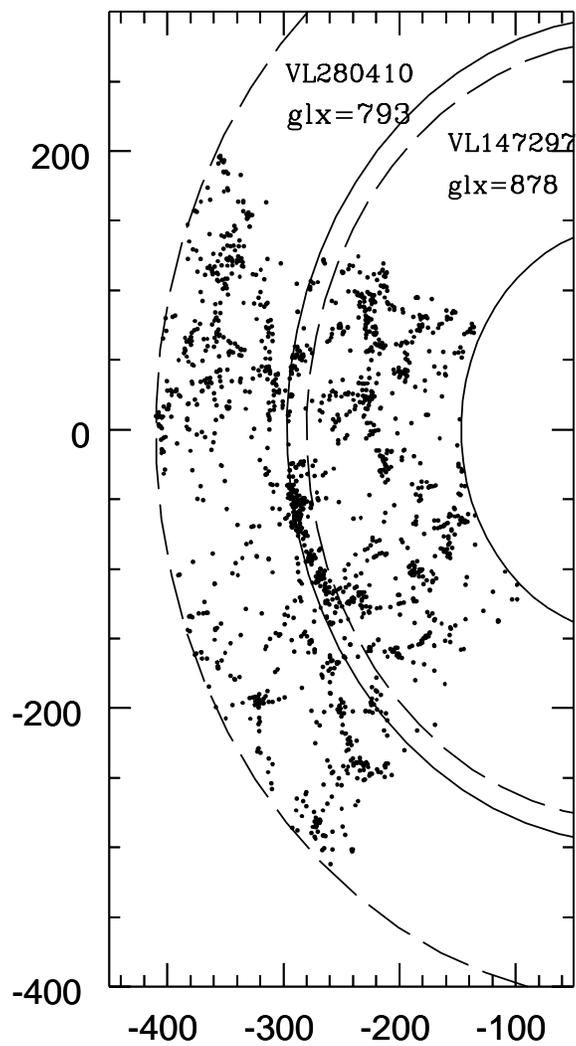
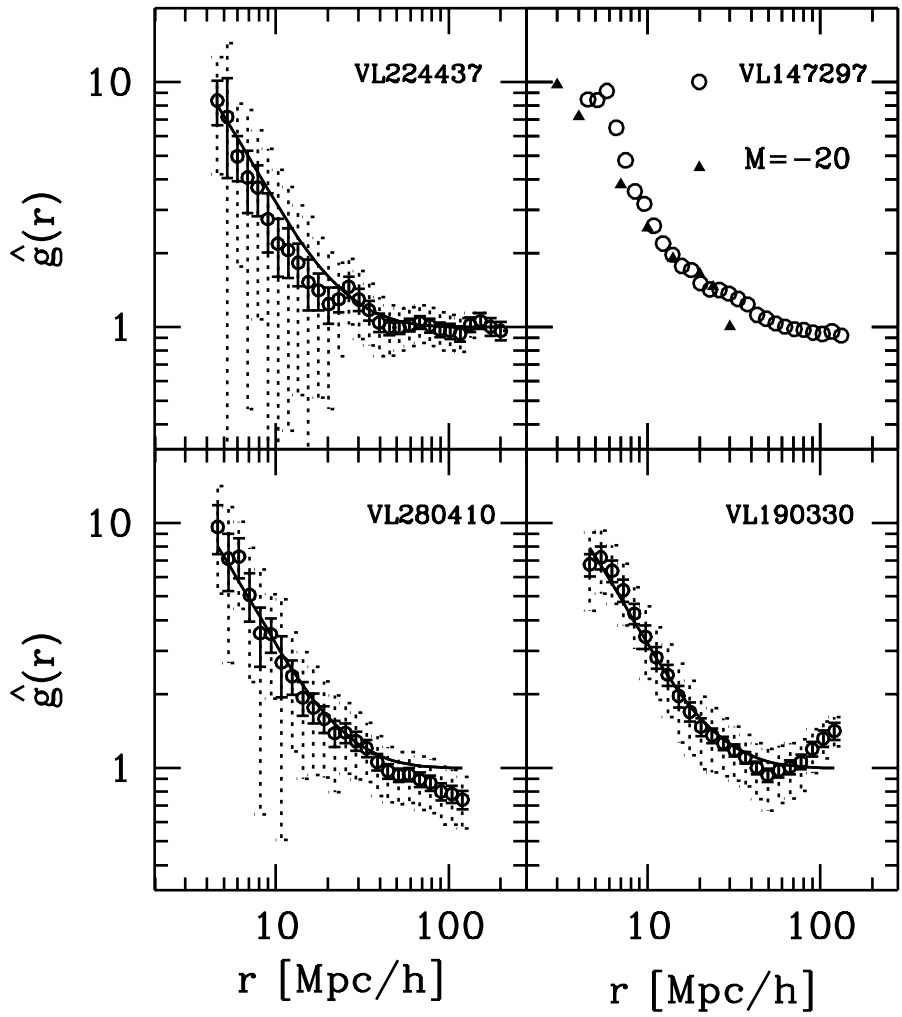
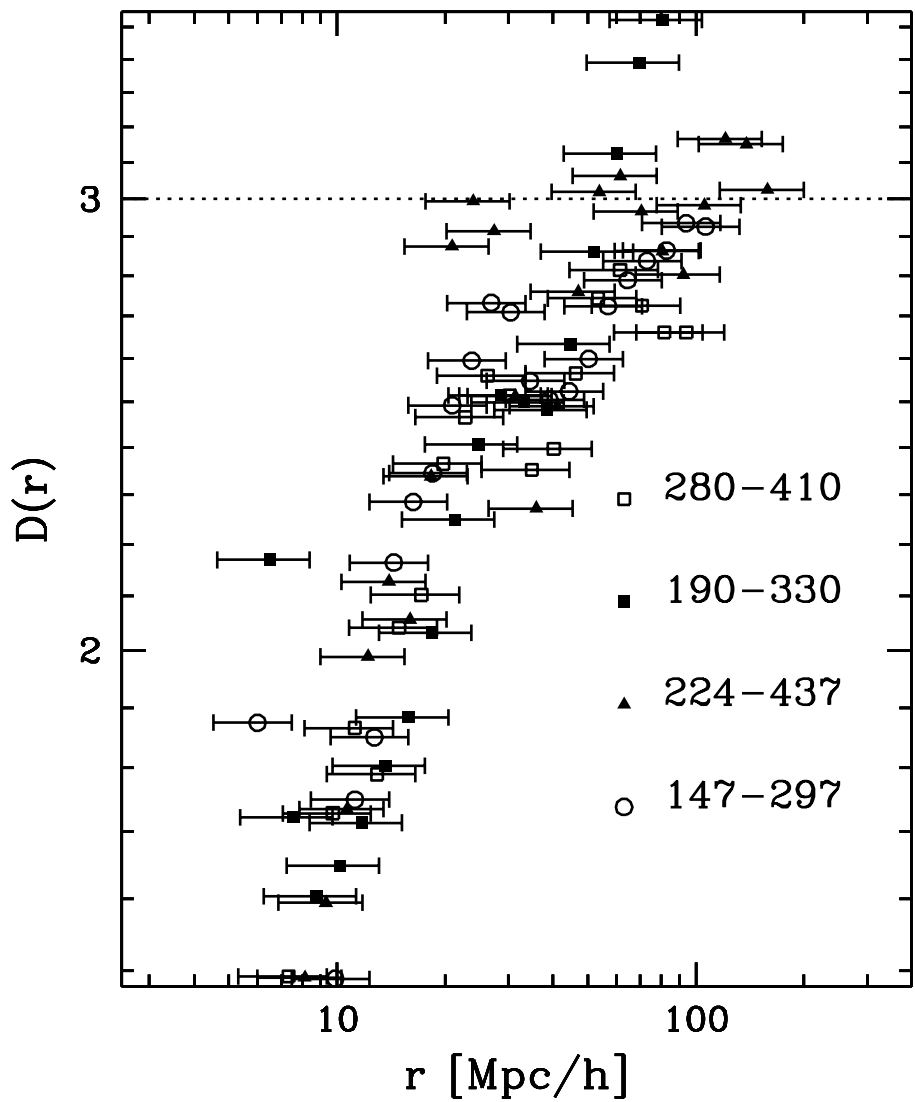
\end{document}